\documentclass{article}
\usepackage[utf8]{inputenc}
\usepackage{xcolor}
\usepackage{authblk}

\title{Exoplanetary Interiors} 
\author[1]{Nadine Nettelmann}
\author[2]{Diana Valencia}
\affil[1]{\small{Institute of Planetary Research, German Aerospace Center (DLR), Berlin, Germany}}
\affil[2]{\small{Centre for Planetary Sciences, University of Toronto, Toronto, ON, M1C 1A4, Canada}}
\date{}
\usepackage{natbib}
\usepackage{graphicx}
\usepackage{geometry}
\usepackage[utf8]{inputenc}
\usepackage{amssymb}

\begin{document}
\maketitle
\newcommand{\Diana}[1]{\textcolor{blue}{{ #1 }}}
\definecolor{gruen}{rgb}{0,0.7,0}
\newcommand{\Nadine}[1]{\textcolor{gruen}{{ #1 }}}

{\centering
\small{To appear as a book chapter in \textit{ExoFrontiers: Big questions in exoplanetary science}}

\footnotesize{accepted 2020/October}}

\section{Introduction} 


With the turn of the century, two discoveries opened new pathways in planetary science. Those were the first mass-estimate of an exoplanet around a Sun-like star, 51 Peg b \citep{MayorQueloz95} through the radial velocity (RV) method and the first radius measurement of an exoplanet, HD209458b \citep{Charbonneau00} through transit light curve (TLC) analysis. These initial discoveries pointed to the challenges of understanding the atmosphere, interior, and evolution of exoplanets including the possibility of mass loss of planets on close-orbits that are exposed to strong irradiation \cite{Guillot96}. They raised the question of heating and inflation mechanisms \cite{GuillotShowman02}, and finally, of the nature of these objects in terms of composition compared to the known planets in the Solar system. The field of exoplanet interior modeling was born.

However the interior of a planet is not directly accessible to observations. The most important observational parameters to infer internal structure refer to their orbits (semi-major axis a, eccentricity $e$), the star (effective temperature $T_{\star}$,  radius $R_{\star}$ yielding the equilibrium temperature $\sigma_B\,T_{eq}^4=(1-A_B)\,T_{\star}^4(a/R_{\star})^2/\sqrt{1-e^2}$ \cite{Miller11}, age $\tau_{\star}$), gravity (planet mass $M_p$, tidal response Love number $k_2$), shape  (radius $R_p$, tidal response $h_2$), as well as atmospheric spectra and phase curves. 
Models must also rely on assumptions. Common assumptions include hydrostatic equilibrium and, for the case for close-in planets, a 1:1 spin-orbit rotational resonance. Planet formation models can provide additional constraints.

\begin{figure}
\includegraphics[width=12cm]{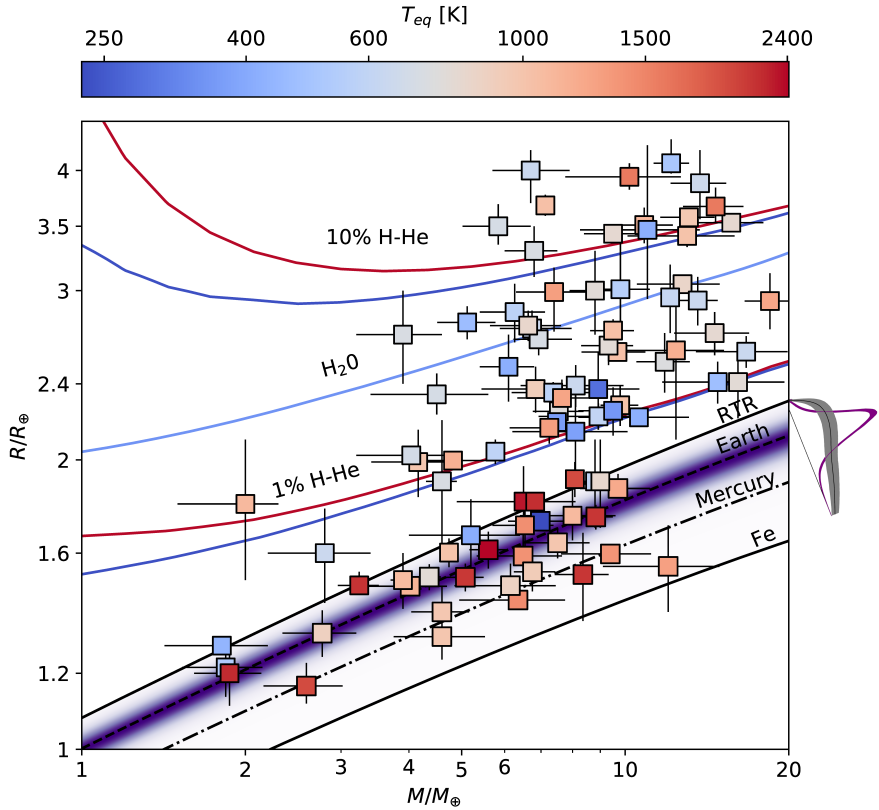}
\caption{\label{fig:MR}
Masses and Radii of supposedly rocky exoplanets with measurement uncertainties  below 25\%, color coded according to $T_{eq}$ (in log-scale). M-R relationships are shown for reference: pure Fe-Ni (core mass fraction cmf=1), Mercury-like (cmf=0.63), Earth (0.325), and the rocky threshold radius (RTR, cmf=0) above which planets have volatiles,  earth-like rock with 1\% and 10\% by mass H-He with equilibrium temperatures of 300K (blue) and 1000K (red), a pure H$_2$O-planet at 1000K equilibrium temperature. Shaded purple region show the M-R of rockky planets with the same refractory ratio as stars (e.g. Fe/Mg and Fe/Si). Outlet figure: Comparison between probability density of stars (purple) and rocky exoplanets (planets within RTR and Fe, grey) in cmf space. 
}
\end{figure}

The first discovered and characterised exoplanets were gaseous planets owing to observational biases, but technological advancements have made discovering smaller planets a routine endeavour, albeit still challenging. In fact, we now know that super-Earths are the most common class of planets on orbital distances out to 500d \citep{Hsu19}, and while the questions regarding interiors are similar to that of gaseous planets, there are also marked differences.  For example, the temperature structure has a negligible effect on the radius of a rocky planet compared to one that has an envelope. We outline big science questions pertaining each class of planets:

\begin{itemize}
\item
{\bf What is the amount of heavy elements in a planet and do all planets possess an iron-rock core?}
Low-mass exoplanets are plagued with compositional degeneracies arising from trade-offs between the different building blocks: iron cores, silicate mantles, water content, and hydrogen+helium gas layers. This means that  interior models constrained by $M_p$ and $R_p$ allow for large spread of  possible compositions.
Famous examples are the warm Neptune GJ436b and the sub-Neptune GJ1214b. Their mass and radius could be explained by a light (few \% $M_p$), dry H/He atmosphere atop a massive rock-iron core but as well by a 90\% (GJ436b) to 100\% (GJ1214b) water interior \citep{Nettelmann10,Nettelmann11,Valencia13,Morley17}. This motivates to search for additional constraints. For GJ436b, formation models were invoked to reduce the degeneracy \citep{Figueira09}; atmospheric spectra suggest a high-Z atmosphere \citep{Morley17} and thus water-rich interior with just a small core. GJ1214b's transmission spectrum is flat due to clouds/hazes \citep{Kreidberg14}, its internal structure remains hidden. An alternative, promising, and novel approach for exoplanets can be measuring their tidal response in form of the Love numbers $h_2$ and $k_2$ \citep{Kramm11,Padovan18}. 

\item
{\bf How much and through what mechanisms are the interiors of planets heated or delayed from cooling?}
Many strongly irradiated gaseous planets require an additional heat source to explain their large radii. Recently, statistical analysis of a large sample of inflated planets in the mass range 0.5--10 $M_J$ has allowed to empirically quantify the amount of extra heating \citep{Thorngren18}. But some individual mysteries exist. An example is the hot Saturn WASP-39b, where even interior models that incorporate heating fall short of explaining the high atmospheric metallicity inferred from transmission spectroscopy \citep{Wakeford18,Poser19}. In that regard, exoplanetary interiors are reaching the level of challenges we encounter in the solar system: state-of-art albeit simple Jupiter models that employ most advanced H/He-equations of state fail to explain observed super-solar atmospheric heavy element abundances \citep{Stevenson20}.

\item
{\bf What is the origin of the observed populations in the radius-period diagram?} 
In the solar system, the outer planets are large, gaseous, and cold while the inner ones are small, rocky, and warm. Planets closer in, at shorter orbital distances, receive stronger irradiation. There is observational indication that these classes, gas-rich vs.~gas-poor, are separated by a gap in radius--irradiation space also in extrasolar systems, albeit for different reasons. Occurrence rates of Kepler planets around  FGK stars revealed a peak  at compact size ($< 1.5\:R_E$) and very high irradiation ($> 100\:S_E$ -- irradiation received by Earth), and a second peak at large radius ($> 2\: R_E$) and low irradiation values ($< 60 \:S_E$) \citep{Fulton17}. The negative slope of the radius--irradiation valley is consistent with compact transiting planets being the result of atmospheric evaporation of gas-born planets \citep{OwenWu13,Lopez13}. However, by extending the analysis to include M dwarf stars, the slope of the valley was found \citep{Cloutier20} to reverse sign making it inconsistent with the mass loss scenario and more compatible with a gas-poor formation environment \citep{Lopez18}.  Objects in and along the radius valley are excellent targets to study planetary formation and evaporation. For a detailed discussion of the radius-period diagram including the radius-gap, the radius desert at very short periods, and the cosmic shoreline separating bodies with and without atmospheres, see Ref.\citep{Zhang20}.

\item 

{\bf What does the composition of rocky planets tell us about their formation?}
The refractory component of low-mass planets should mirror that of the host star \citep{Dorn17, Hinkel18}, thus setting the rock to iron core ratio to that of the star and reducing the degeneracy by one degree of freedom.  However, the data on rocky planets do not necessarily support this idea \citep{Plotnykov20}. In addition, planets more iron-rich than Mercury (e.g. Super-Mercuries) are found \citep{Plotnykov20}, as well as planets that if rocky, are depleted in iron with respect to Earth (e.g. Super-Moons).  We do not have a reliable formation theory that explains their existence \citep{Scora20}. Figure \ref{fig:MR} shows the current known exoplanets with measured mass and radius with an error below 25\%. The rocky threshold line shows the maximum size a rocky planet has including all the relevant Mg-Si-O mineralogy. The curve for pure Fe$_{0.95}$-Ni$_{0.05}$ shows an extreme composition below which planets are expected to not exist. Any rocky planet is expected to lie between these lines. To compare to the stars' refractory composition,  Ref.~\citep{Plotnykov20} translated the stars refractory ratios into $M-R$ relationships (red). Apparently, the detected planets span a wider distribution that needs to be explained by any successful formation theory.
\end{itemize}

\section{Important questions and goals}\label{sec:nn_giants}

\paragraph{What are the structure parameters we are interested in?}
For gas giants, the information we aim to obtain is foremost the bulk mass fraction of heavy elements $Z_p$ (elements that are heavier than hydrogen and helium), and the corresponding gas mass fraction (assuming H and He in roughly protosolar mass ratio of $\rm M_{He}/(M_{He}+M_H)=0.270$). A planet's $Z_p$ value yields a first classification as a terrestrial or super-Earth ($Z=1$) , sub-Neptune (high $Z_p$, but also with gaseous H/He and/or water), ice giant (high $Z_{H2O} > 0.5$ is a solution), or gas giant planet ($Z_p < 0.5$).  In practice, theoretical mass-radius relations for different assumed compositions serve to invert an observed $M_p$--$R_p$ pair for possible composition. For an extensive set of $M_p$--$R_p$ relations for gaseous planets ranging from $M_p=10\:M_E$ to 10 $M_J$ see \citep{Fortney07}, for rocky and water-rich planets in the 0.4--20 $M_E$ mass range and including thermal effects in the water EoS see \citep{Valencia13,Madhu16}, and with $S_E$ as a third parameter see \citep{Zeng19}.

Second, we are interested in the distribution of heavy elements in the interior, such as the presence of a core. 
In the case of gas giants we usually refer to the core as the refractory material that can either be at the centre, or diluted, meaning it is mixed into the envelope up to some radius but with high-Z at the center. In the case of terrestrial bodies, the core refers to the iron-rich central mass agglomeration below a silicate mantle.  For gaseous planets, simple two-parameter models would quantify the distribution as $Z_{env}$ and $M_{core}$, with $Z_p = Z_{env}(M_p-M_{core}) + M_{core}$. Mass and radius alone are not enough to provide $M_{core}$ even if they would be exactly known, because $R_p(M_p)$ is only one constraining parameter. Uncertainties in the temperature profile, which underneath the radiative atmosphere is assumed to be adiabatic but the transition point is model-dependent \cite{ThorngrenGao19}, adds to the uncertainty in $Z_p$. 
The high observed heat flow out of Jupiter indeed suggests a convective interior with adiabatically stratified temperature profile, although Jupiter and Saturn may have super-adiabatic layers. The assumption of an adiabatic interior without extra heating (beyond insolation) allows us to estimate the minimum $Z_p$ value of a giant planet.

\paragraph{What is the composition of the envelope of sub-Neptunes?}
Sub-Neptunes are small planets that have a gaseous envelope. It may be light (few \% $M_p$), but large enough to affect the planetary radius. This is in contrast to rocky planets, like Earth and Venus, that have an atmosphere that does not affect the radius in a considerable way.   The most important science goal for sub-Neptunes is to determine what the composition of the envelope is, as it holds key information to their formation.  Conventional formation theories predict that planets that form outside the snowline will have large water contents. These may or may not migrate to the close orbits. Instead,  if formed in situ, close-in sub-Neptune planets should have negligible water. The problem is that from $M-R$ pair data it is not possible to discern between either scenario (water-rich vs,~water-poor). And even with good atmospheric charaterisation, we would need to connect the compositon of the upper millibars of atmosphere to that of the bulk envelope.

\paragraph{What is the origin of iron-poor or iron-rich rocky planet?}
Planets that are sufficiently heated and compact enough to be well within the rocky region, should have no water and no H-He. This is because a steam atmosphere or any H-He enlarges the planet's radius considerably.  Thus, compact and highly irradiated planets are presumed to be  completely rocky, composed of silicate mantles and iron cores. Ref.~\citep{Plotnykov20} constrained the composition of $\sim30$ rocky planets and obtained the refractory ratios Fe/Mg and Fe/Si. Compared to planet-hosting stars, rocky planets span a wider range in composition, see Figure \ref{fig:MR}. They find a prevalence of super-Mercuries in the exoplanet data, as well as planets that are depleted in iron (super-Moons) should they truly be rocky.  At the moment we lack a reliable theory that explains both iron enriched and iron-depleted massive rocky planets \citep{Scora20}.  Alternatively, the planets classified as iron-depleted instead could have very small amounts of volatiles.  However, some of them are also highly irradiated ($\gtrsim$ 300 $S_E$) around FGK stars, and thus, expected to have lost their atmospheres.   Either way, these planets are a mystery.  Perhaps they have water-rich atmospheres that have undergone runaway greenhouse states \citep{Turbet19}, which is the predicted fate of Earth in about 600 Myr due to the evolution of the Sun.

\paragraph{We expect inferred metallicities from structure models to agree with those from planet formation models.}

\begin{figure}
\includegraphics[width=12cm]{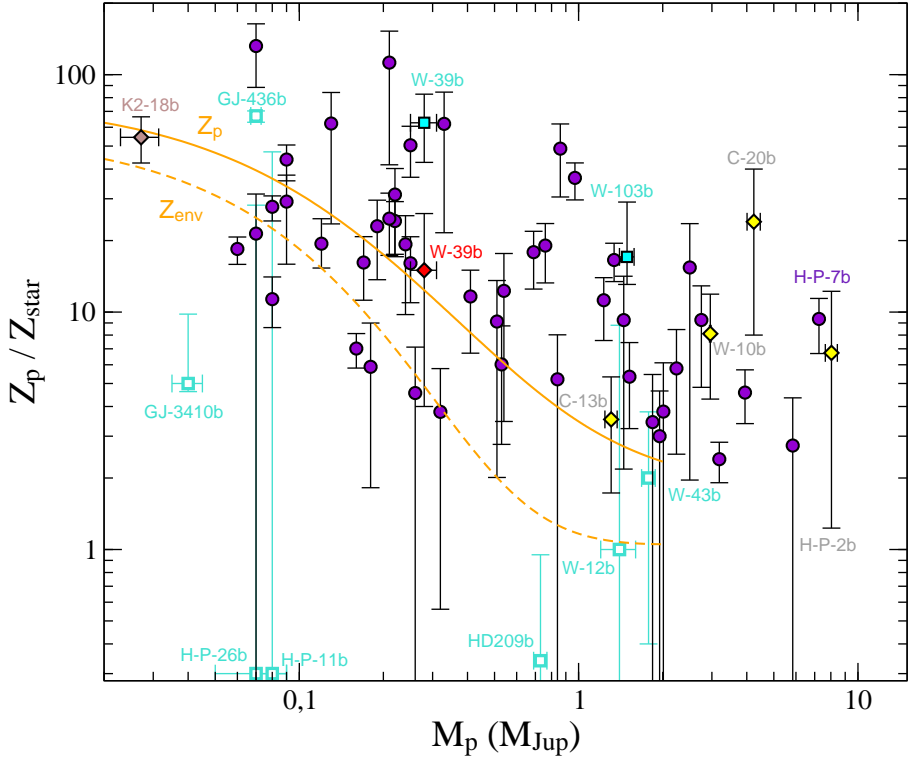}
\caption{\label{fig:nn_zp_zstar}
Compilation of heavy element enrichment by mass over planet mass; \emph{violet, circles}: bulk $Z_p$ for the sample of 47 non-inflated planets in Ref.~\citep{Thorngren16}; \emph{yellow diamonds}: selection of inflated planets which are predicted to carry a large amount of heavy elements even though extra heating is not included in the interior model calculation; \emph{red diamond}: WASP-39b based on models that account for extra heating and delay of cooling by clouds \citep{Poser19}; \emph{brown diamond}: super-Earth K2-18b \citep{Madhu20}; \emph{Orange curves} show fits to the prediction for $Z_{env}$ and $Z_p$ from CA formation \citep{Venturini16}. \emph{Cyan squares} are atmospheric metallicities,  \emph{open}: from Ref.~\citep{Chachan19} with observed values for HAT-P-11b and HAT-26b lower than shown here (placed at 0.2) and likely influenced by clouds; \emph{filled}: WASP-39b \citep{Wakeford18} and WASP-103b \citep{Kreidberg18}.} 
\end{figure}

An assumption for transiting gaseous exoplanets is that they formed somewhere in the disk, migrated inward, and finalized mass accretion a few Myr after the young star reached the main sequence. Their inferred composition today should then be in agreement with predictions from planet formation models.

Core accretion formation (CA) assumes the formation of initial rocky core seeds from dust in the protostellar disk. The core then grows with some gas accretion. Captured planetesimals may dissolve in the gaseous envelope or sink to the core. Once the envelope mass equals about the core mass, which can occur for $M_{core}=1$ to several 10 $M_E$, run-away gas accretion sets in. Mass of the core, bulk $Z_p$, and final mass are thereby dependent also on gas opacity and solids surface density. In any event, CA formation predicts an inhomogeneous structure with a slightly super-stellar heavy element abundance in the gaseous envelope.  Jupiter is an example \citep{Vazan18}. In the CA model, and also in the gravitational disk instability model (GI), planets the mass of Jupiter and higher are massive because they accreted so much gas from the disk. $Z_p$ should decrease the more massive the gaseous envelope is \citep{Venturini16}. 

The  $Z_p$ value inferred from planetary interior modeling depends on internal temperatures. Suppose two planets of same mass and age, but one inflated and the other one not. The inflated one will be larger and warmer since an inflated planet experiences mechanisms that heat the interior (tidal, Ohmic) or delay its cooling (high opacity in the atmosphere). As a result, including extra heating/delay of cooling in the modeling and evolution runs leads to higher inferred $Z_p$ values.

In Figure \ref{fig:nn_zp_zstar} we plot the inferred heavy element enrichment of the sample of 47 non-inflated planets ($T_{eq}<1000$~K) with measured mass and radius known by 2016 according to the models in Ref.~\citep{Thorngren16}. The predicted trend from CA formation of $Z_p$ decreasing with $M_p$ can clearly be seen. However, $Z_p(M_p)$ of planets more massive than Jupiter does not approach the stellar values $Z_{\rm star}$ toward the brown dwarf regime ($M_{\rm BD}=13$--$80 M_J$) as predicted by both CA and GI formation: the heavy element enrichment of massive planets tends to be higher. For the $\sim 7 M_J$ planet HAT-P-20b, $Z_p\sim 9\times Z_{\rm star}$ implies the enormous amount of heavy elements of 550--750$\,M_E$ \citep{Thorngren16}. This is 7--9 times the mass of heavy elements confined to the planets of the solar system. There are also a couple of inflated and massive hot Jupiters for which models predict high enrichment even though extra heating has not even been taken into account (yellow diamonds). Huge amounts of hundreds of $M_E$ of heavy elements in a single planet are not predicted by any formation theory. 

We also plot in Figure \ref{fig:nn_zp_zstar} atmospheric metallicities ($Z_{atm}$) obtained from transmission and emission spectra \citep{Wakeford18,Kreidberg19,Chachan19}. Generally one expects $Z_{atm} \leq Z_p$ since $Z_{atm} > Z_p$
would trigger an instability that erases the inversion. The finding $Z_{atm}>Z_p$ for WASP-39b is a mystery.

\paragraph{Tidal response can be used to infer internal density distribution.}

In order to constrain both $M_{core}$ and $Z_{env}$, or for terrestrial planets the iron-core mass fraction, a further gravitational parameter is required. This may be the tidal Love number $k_2$ or the shape parameter $h_2$. In linear approximation, $h_2=1+k_2$ \cite{Love11}. 

A close-in planet suffers  tidal deformation. In 1:1 spin-orbit resonance and neglecting librations around the equilibrium position, the tidal bulge of the planet is permanently elongated toward the star; the response is static. One can define a linear response coefficient for the tidal shape deformation with respect to the unperturbed shape, $h_2$, and an equivalent coefficient for the gravity field perturbation, $k_2:= - V_{22}/W_{22}$, 
where $V_{22}$ and $W_{22}$ are (tesseral) harmonics of the planetary and stellar gravity field evaluated at the planet's surface, respectively \citep{Nettelmann19}. 

For fluid planets and relaxed solid planets, $k_2$ is dictated by the internal density distribution. By measuring $k_2$ (or $h_2$) an upper limit can be placed on the mass of the central core. For instance, if the sub-Neptune GJ1214b were a water world it would have $k_2\sim~0.9$, whereas $k_2<0.1$ if it were composed of a rocky core with H/He atmosphere \citep{Nettelmann11}.  For predictions of $k_2$ values in the fluid regime of super-Earths see \citep{Kellermann18} and for the warm Neptune GJ$\,$436b see \citep{Kramm11,Padovan18}.

\begin{figure}
\includegraphics[width=12cm]{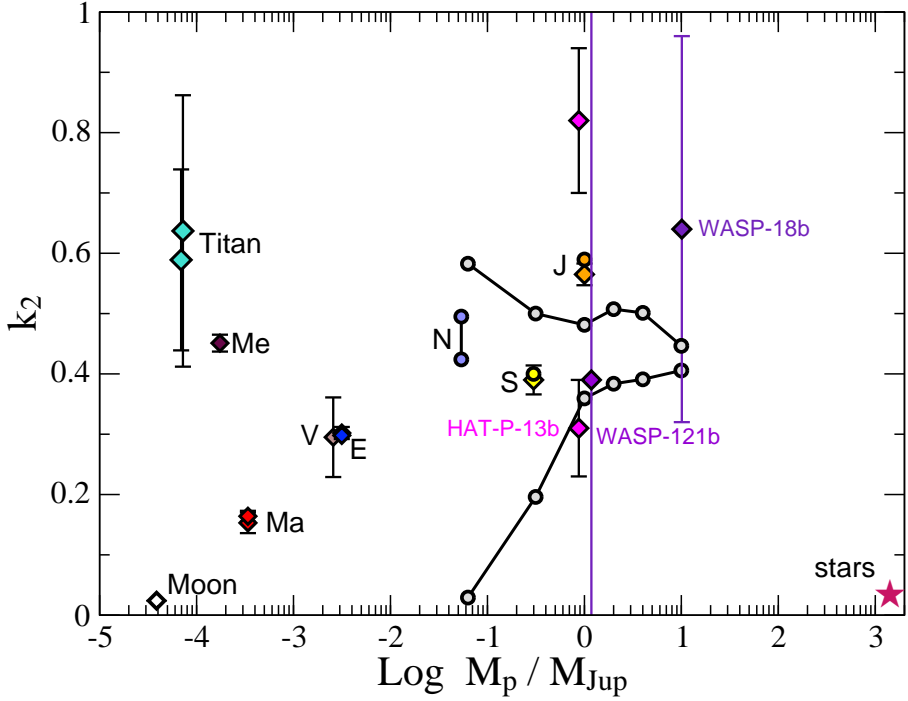}
\caption{\label{fig:nn_k2}
Observed Love number $k_2$ values \emph{(diamonds)} in the solar system and of three exoplanets as labeled. The value for WASP-121b was derived from its $h_2$ estimate \cite{Hellard20}, for WASP-18b from RV variations \cite{Csizmadia19}, for HAT-13b assuming locked apsidal motion between planets b and c \citep{Buhler16,Hardy17}, for Saturn from Cassini and groundbased long-term astrometric data \citep{Lainey17}, and Jupiter's value is the current Juno observation \cite{Durante20}. \emph{Circles} show calculated values from interior models assuming static tidal response; the value for Saturn is for a rotation rate of 10h32m min, for Neptune for respectively $16^h 06^m 40^s$ and $17^h 27^m 29^s$ \cite{Helled10}; the two curves with \emph{gray circles} are for irradiated exoplanets with $Z_p(M_p)$ relation after \cite{Thorngren16}, where the upper curve assumes homogeneous distribution (small $M_{core}=1$ $M_E$, high $Z_{env}$) and the lower curve heavy elements to be confined to the core (low $Z_{env}=1x$ solar, high $M_{core}$). Since thermal structure matters, these exoplanet models have been evolved to 4--6 Gyr.
}
\end{figure}

Figure \ref{fig:nn_k2} shows $k_2$ values for planets and satellites. Among terrestrial bodies, the spread in $k_2$ is large  as a result of different internal density distributions, rheologies (Moon vs. Mercury), and possible internal oceans (Titan). For fluid exoplanets we highlight the potential of $k_2$ to reveal internal density distribution with the help of the two black curves. Both curves are for planets of same mass, same orbital distance of 0.1 AU around the Sun, same age of about 4--6 Gyr, and notably, of same composition $Z_p(M_p)$ following roughly the exoplanet distribution in Figure \ref{fig:nn_zp_zstar}. The upper curve is for homogeneous irradiated planets with $M_{core}=1$ and $Z_{env}=0.78$ at $M_p=20$ (Neptune-like), $Z_{env}=0.3$ at $M_p=100\:M_{\rm E}$ (Saturn-like), $Z_{env}=0.1$ at $M_p=1$--$10\: M_{\rm J}$ (gas giants). Models along the lower have curve that same amount of heavy elements but confined to the core ($Z_{env}=0.01$), leading to a strong reduction in $k_2$ especially for the particularly degenerate class of (sub-)Neptune sized planets. Even the curve for a homogeneous planet underestimates the value of Jupiter, which has some central mass condensation: this is because Jupiter's atmosphere is cold and dense while that of an irradiated hot Jupiter is warm and more extended. Both bracketing curves fail to explain the high $k_2$ value of $\sim 0.60$ of WASP-18b \citep{Csizmadia19}: this is a puzzle.

\section{Challenges}  

\paragraph{Formation and evolution of super-Mercuries and super-Moons} 
A large compositional diversity in presumed rocky planets has been found \citep{Plotnykov20}, larger than the Fe/Si of the planet-hosting stars. This diversity includes planets more iron-rich than Mercury (e.g. Super-Mercuries) that are difficult to form even by invoking giant planet collisions of differentiated massive embryos \citep{Scora20}. Also, if truly rocky, there is subsample of planets 2-fold depleted in Fe/Mg with respect to stars.  It is essential that we determine if these are truly bare rocks because at the moment we have no theory that predicts preferential loss of iron compared to Mg+Si rocks \citep{Scora20}. One avenue is to look at the heat redistribution of their phase curves to rule out the presence of atmospheres \citep{Kreidberg19}.    

\paragraph{One-to-One Stellar-Planet Compositional Comparison}
There is tremendous potential to learn about planet formation in systems where both the planets refractory composition and the host stars is known. At the moment, there is only a handful of planets for which this is the case \citep{Plotnykov20}. Unfortunately, the error bars in planetary radius, but especially in planetary mass, as well as in stellar refractory ratios are too large to yield definite conclusions.  It will be even more powerful to compare multiplanet systems, given the difference in planetary irradiation.

\paragraph{Love number computation and measurements}
In the fluid regime, the Love numbers $k_2$ and $h_2$ can straightforwardly be computed from the internal density distribution. However, the relaxation time  of solid planets or solid rocky cores into the fluid regime depends on the rheology (viscosity, rigidity) of the material and the frequency of tidal forcing. Mars and Earths are not in the fluid regime.  Viscosity of solid materials and the unknown rotation rate are a major source of uncertainty for $k_2$ of solid planets. For Jupiter, Juno spacecraft measurement just revealed a deviation from the theoretical static response value of 0.5897 \citep{Nettelmann19,Wahl20} by $-2\%$ \citep{Durante20}, teaching us that dynamic effects play a role, which are much more challenging to quantify.

Current measurement uncertainties for exoplanets are large and at present preclude a robust estimate of core mass upper limit. For TLC method, uncertainties in $h_2$ are predicted to be at best $\pm 20\%$ with upcoming instrumentation (PLATO, JWST) \citep{Hellard20}. Nevertheless, long-term observations in this new field will help to reduce uncertainties.

\paragraph{Internal Temperature profile}
In  99\% of Jupiter's mass, hydrogen is in a metallic, quantum-mecha\-ni\-cally degenerate state. The opacity is high and thermal conductivity hampered by moderate temperatures ($<20,000$~K) and high densities ($\sim 1$ g/cc). Similar conditions are predicted to occur in more massive planets and brown dwarfs \citep{Becker18}, suggesting a largely adiabatic interior. 
However, observed oscillations in the C-ring of Saturn's indicate a large stably stratified core \citep{Fuller14}. If super-adiabatic, inferred metallicities could be significantly higher, and even explain some inflated planets \citep{Chabrier07}. Uncertainties in thermal structure and thus inferred metallicity could be large for sub-Neptune to Saturn-sized exoplanets, even in the absence of extra heating, rendering inferring their 'nature' challenging. 

Furthermore, the temperature structure of a planet has 
a profound effect on the state of the matter and related properties. The presence of liquid iron cores that may produce magnetic fields, surface magma oceans that can create specific emission signatures, or partially molten mantles that can affect the outgassing, and tidal dissipation of the planet, all depend on the mode of heat transport and temperature structure. 

\paragraph{Equations of State (EoS)  and phase diagrams}
Interior structure models rely on adequate equations of state for the end-members (H-He, H$_2$O/ices, rocks, metallic iron) and their mixtures and phase diagrams. Uncertainties in melting lines, for instance, add to the uncertainty of whether super-Earths possess a magnetic field from a dynamo that operates in a partially molten silicate layer or in a liquid iron core. While the EoS and phase diagram of pure water under planetary interior conditions is well understood (uncertainties still persist regarding lattice structures in super-ionic and ice phases at high pressures outside the realm of planetary interiors), solubility of water with methane, ammonia, and refractory materials is not well known. Ab initio calculations \citep{Wahl13} predict  solubility of water, iron, and also of Si and Mg at least at low concentrations of 1:256 in metallic hydrogen, which occurs at about $P>0.5$ Mbar in H-rich planets. This suggests that planets more massive than Neptune, where pressures of 0.5 Mbar are reached at $\sim 0.7 R_{Nep}$, have partially mixed interiors while low-mass planets may harbour segregated rock-iron cores. Of importance to habitability is the melting lines of rock material, (as it connects to outgassing ad tidal dissipation), and iron alloys (as it relates to magnetic field generation). More experimental and theoretical work is needed to understand the mixing behavior of multi-component planetary materials.



\paragraph{Inflation mechanisms}
Mechanisms that tranfer energy from the star to the planet  not well understood,  in particular how much and where in the planet the heat is deposited as a function of $M_p$, and $T_eq$. Tidal heating (due to eccentricity and/or obliquity) goes with $e^2/Q_p$ offering the opportunity to estimate a planet's tidal quality factor $Q_p$. For gas giants, several studies come to the conclusion that the efficiency of Ohmic heating peaks at $T_{eq} \sim 1500$--1600 K \citep{Thorngren18}; Ohmic heating may also be relevant for sub-Neptunes \citep{PuValencia17} but has not systematically been studied for Neptune-to-Saturn mass planets.

\section{Opportunities} 

\paragraph{New observational facilities}
 At the moment the TESS space telescope is surveying the closest M Dwarfs and measuring the radii of super-Earths and large exoplanets.  The next leap into observation capabilities will come with the JWST scheduled to launch late 2021. This telescope is expected to revolutionize our understanding of the atmospheres of exoplanets by routinely obtaining transit spectroscopy for the exoJupiters, and sub-Neptunes. PLATO (launch date 2026) will survey more than 1 million stars and its focus is to find rocky planets in the habitable zone of Sun-like Stars. Following, Ariel will launch in mid 2028 and will also obtain the composition and thermal structure of atmospheres, but contrary to JWST, it can dedicate more time for each planet observed. All these space missions are being supported by existing and in-construction ground facilities that measure planetary masses.  With all these resources, the future of exoplanets research is bright.

\paragraph{Exoplanets with liquid/icy water}

Due to observational biases, exoplanets with liquid or icy water on the surface have not been identified yet. However, soon we expect detections of compact long-period planets around M Dwarfs, some of which could harbor oceans.
The high $k_2$ value of Saturn's moon Titan (Figure \ref{fig:nn_k2}) is indicative of a deep water ocean, and ESA's JUICE mission (launch 2022) and NASA's Europa Clipper (launch 2024) will set out to measure the depths of sub-surface oceans on Ganymede and Europa by means of tidal distortion. Measuring the tidal response of distant low-mass planets is beyond the scope of near-future instrumentation and thus offers opportunities for future generations of exoplanetary scientists.

\paragraph{Love number measurements}
Within just a decade, four methods (TLC, RVV, TTV, orbital fixed points) have been elaborated to observationally infer $k_2$ or $h_2$. While uncertainties are still large for each single method application of these independent methods to the same planet could greatly reduce the uncertainty, like the observable part of the universe is now known to be globally flat ($\Omega=1$) due to independent observations of Type 1 Super Novas, of the cosmic microwave background, and of baryon oscillations \citep{Frieman08}. The RVV method most benefits from longer baseline observations (decades instead of years), and the TLC method from long exposure times for single planets \citep{Hellard20}.

Since assumed static tidal response of afluid body places a strict upper limit on the $k_2$ value, higher observed values offer the unique opportunity to study dynamic effects. Future extended RV observations of the ultra-hot Jupiter WASP-18b \citep{Csizmadia19}, which has maximum static $k_2$ of only 0.43 may reveal if that era has already begun.

\paragraph{Statistical analysis of increasingly large samples}
As the sample size of exoplanets has grown, so has the ability to extract information from the planets as a population. Statistical studies have aimed at uncovering any trends in the data: planetary mass and star metallicity \citep{Miller11,Mulders18, Neves12}, radius and irradiation \citep{Thorngren18}, composition and irradiation \citep{Plotnykov20}, etc. This trend will continue to grow, especially for planets around M dwarfs and evolved stars, and small planets in general, as the observational challenges are being overcome.  There is power in numbers.  Even if not as precise, sophisticated statistical analysis can yield quantitative constraints on planet formation and evolution if the sample size is large and informative enough. In parallel there is a shortcut to planetary understanding by focusing on the outliers that challenge conventional wisdom.



\bibliographystyle{plain}

\begin{thebibliography}{10}

\bibitem{Becker18}
A.~Becker, M.~Bethkenhagen, C.~Kellermann, and R.~Redmer.
\newblock {Material Properties for the interiors of massive GPs and BDs}.
\newblock {\em AJ}, 156:4, 2018.

\bibitem{Buhler16}
P.B. Buhler, H.A. Knutson, K.~Batygin, B.~Fulton, J.J. Fortney, A.~Burrows, and
  I.~Wong.
\newblock {Dynamical constraints on the core mass of hot Jupiter HAT-P-13b}.
\newblock {\em ApJ}, 821:26, 2016.

\bibitem{Chabrier07}
G.~Chabrier and I.~Baraffe.
\newblock {Heat transport in (exo)planets: a new perspective}.
\newblock {\em ApJ}, 661:L81, 2007.

\bibitem{Chachan19}
Y.~{Chachan}, H.A. {Knutson}, P.~{Gao}, T.~{Kataria}, I.~{Wong}, G.W. {Henry},
  B.~{Benneke}, M.~{Zhang}, J.~{Barstow}, J.L. {Bean}, T.~{Mikal-Evans}, N.K.
  {Lewis}, M.~{Mansfield}, M.~{L{\'o}pez-Morales}, N.~{Nikolov}, D.K. {Sing},
  and H.~{Wakeford}.
\newblock {A Hubble PanCET Study of HAT-P-11b: A Cloudy Neptune with a Low
  Atmospheric Metallicity}.
\newblock {\em AJ}, 158:244, 2019.

\bibitem{Charbonneau00}
D.~Charbonneau, T.M. Brown, D.W. Latham, and M.~Mayor.
\newblock {Detection of planetary transits across a sun-like star}.
\newblock {\em ApJ}, 529, 2000.

\bibitem{Cloutier20}
R.~{Cloutier} and K.~{Menou}.
\newblock {Evolution of the Radius Valley around Low-mass Stars from Kepler and
  K2}.
\newblock {\em AJ}, 159:211, 2020.

\bibitem{Csizmadia19}
{Sz}. Csizmadia, H.~Hellard, and A.M.S. Smith.
\newblock {An estimate of the $k_2$ Love number of WASP-18Ab from its radial
  velocity measurements}.
\newblock {\em A\& A}, 623:A45, 2019.

\bibitem{Dorn17}
C.~{Dorn}, N.R. {Hinkel}, and J.~{Venturini}.
\newblock {Bayesian analysis of interiors of HD 219134b, Kepler-10b,
  Kepler-93b, CoRoT-7b, 55 Cnc e, and HD 97658b using stellar abundance
  proxies}.
\newblock {\em A{\&}A}, 597:A38, 2017.

\bibitem{Durante20}
D.~{Durante}, M.~{Parisi}, D.~{Serra}, M.~{Zannoni}, V.~{Notaro},
  P.~{Racioppa}, D.~R. {Buccino}, G.~{Lari}, L.~{Gomez Casajus}, L.~{Iess},
  W.~M. {Folkner}, G.~{Tommei}, P.~{Tortora}, and S.~J. {Bolton}.
\newblock {Jupiter's Gravity Field Halfway Through the Juno Mission}.
\newblock {\em GRL}, 47:e86572, 2020.

\bibitem{Figueira09}
P.~Figueira, F.~Pont, C.~Mordasini, Yann Alibert, C.~Gorgy, and W.~Benz.
\newblock {Bulk composition of the transiting hot Neptune around GJ 436}.
\newblock {\em A\&A}, 493:671, 2009.

\bibitem{Fortney07}
J.J. Fortney, M.S. Marley, and J.W. Barnes.
\newblock {Planetary radii across five orders of magnitude in mass and stellar
  insolation: Application to Transits}.
\newblock {\em ApJ}, 659:1661, 2007.

\bibitem{Frieman08}
J.A. {Frieman}, M.S. {Turner}, and D.~{Huterer}.
\newblock {Dark energy and the accelerating universe.}
\newblock {\em Ann.~Rev.~Astron.~Astrophys.}, 46:385--432, 2008.

\bibitem{Fuller14}
J.~{Fuller}.
\newblock {Saturn ring seismology: Evidence for stable stratification in the
  deep interior of Saturn}.
\newblock {\em Icarus}, 242:283--296, 2014.

\bibitem{Fulton17}
B.J. {Fulton}, E.A. {Petigura}, A.W. {Howard}, H.~{Isaacson}, G.W. {Marcy},
  Phillip~A. {Cargile}, L.~{Hebb}, L.M. {Weiss}, J.A. {Johnson}, T.D. {Morton},
  E.~{Sinukoff}, I.J.M. {Crossfield}, and Lea~A. {Hirsch}.
\newblock {The California-Kepler Survey. III. A Gap in the Radius Distribution
  of Small Planets}.
\newblock {\em AJ}, 154:109, 2017.

\bibitem{Guillot96}
T.~Guillot, A.~Burrows, W.B. Hubbard, J.I. Lunine, and D.~Saumon.
\newblock {Giant planets at small orbital distances}.
\newblock {\em ApJ}, 459:L35, 1996.

\bibitem{GuillotShowman02}
T.~Guillot and A.P. Showman.
\newblock {Evolution of ''51 Pegasus b-like'' planets}.
\newblock {\em A\&A}, 385:156, 2002.

\bibitem{Hardy17}
R.A. Hardy, J.~Harrington, M.~Hardin, M.~Madhusudhan, T.J. Loredo, R.C.
  Challener, A.S. Foster, P.E. Cubillos, and J.~Blecic.
\newblock {Secondary Eclipses of HAT-P-13b}.
\newblock {\em ApJ}, 836:143, 2017.

\bibitem{Hellard20}
H.~{Hellard}, Sz. {Csizmadia}, S.~{Padovan}, F.~{Sohl}, and H.~{Rauer}.
\newblock {HST/STIS Capability for Love Number Measurement of WASP-121b}.
\newblock {\em ApJ}, 889:66, 2020.

\bibitem{Helled10}
R.~Helled, J.D. Anderson, and G.~Schubert.
\newblock {Uranus and Neptune: shape and rotation}.
\newblock {\em Icarus}, 210:446, 2010.

\bibitem{Hinkel18}
N.R. {Hinkel} and C.T. {Unterborn}.
\newblock {The Star-Planet Connection. I. Using Stellar Composition to
  Observationally Constrain Planetary Mineralogy for the 10 Closest Stars}.
\newblock {\em ApJ}, 853:83, 2018.

\bibitem{Hsu19}
D.C. {Hsu}, E.B. {Ford}, D.~{Ragozzine}, and K.~{Ashby}.
\newblock {Occurrence Rates of Planets Orbiting FGK Stars: Combining Kepler
  DR25, Gaia DR2, and Bayesian Inference}.
\newblock {\em AJ}, 158:109, 2019.

\bibitem{Kellermann18}
C.~Kellermann, A.~Becker, and R.~Redmer.
\newblock {Interior structure models and fluid Love numbers of exoplanets in
  the super-Earth regime}.
\newblock {\em A\&A}, 615:A39, 2018.

\bibitem{Kramm11}
U.~Kramm, N.~Nettelmann, R.~Redmer, and D.S. Stevenson.
\newblock {On the degeneracy of the tidal Love number $k_2$ in multi-layer
  planetary models}.
\newblock {\em A\&A}, 528:A18, 2011.

\bibitem{Kreidberg14}
L.~{Kreidberg}, J.L. {Bean}, J.-M. {D{\'e}sert}, M.R. {Line}, J.J. {Fortney},
  N.~{Madhusudhan}, K.B. {Stevenson}, A.P. {Showman}, D.~{Charbonneau}, P.R.
  {McCullough}, S.~{Seager}, A.~{Burrows}, G.W. {Henry}, M.~{Williamson},
  T.~{Kataria}, and D.~{Homeier}.
\newblock {A Precise Water Abundance Measurement for the Hot Jupiter WASP-43b}.
\newblock {\em ApJL}, 793:L27, 2014.

\bibitem{Kreidberg19}
L.~{Kreidberg}, D.D.B. {Koll}, C.~{Morley}, R.~{Hu}, L.~{Schaefer},
  D.~{Deming}, K.B. {Stevenson}, J.~{Dittmann}, A.~{Vanderburg}, D.~{Berardo},
  X.~{Guo}, K.~{Stassun}, I.~{Crossfield}, D.~{Charbonneau}, D.W. {Latham},
  A.~{Loeb}, G.~{Ricker}, S.~{Seager}, and R.~{Vanderspek}.
\newblock {Absence of a thick atmosphere on the terrestrial exoplanet LHS
  3844b}.
\newblock {\em Nature}, 573:87--90, 2019.

\bibitem{Kreidberg18}
L.~{Kreidberg}, M.R. {Line}, V.~{Parmentier}, K.B. {Stevenson}, T.~{Louden},
  M.~{Bonnefoy}, J.K. {Faherty}, G.W. {Henry}, M.H. {Williamson}, K.~{Stassun},
  T.G. {Beatty}, J.L. {Bean}, J.J. {Fortney}, A.P. {Showman}, J.-M.
  {D{\'e}sert}, and J.~{Arcangeli}.
\newblock {Global Climate and Atmospheric Composition of the Ultra-hot Jupiter
  WASP-103b from HST and Spitzer Phase Curve Observations}.
\newblock {\em AJ}, 156:17, 2018.

\bibitem{Lainey17}
V.~Lainey, R.A. Jacobson, R.~Tajeddine, N.J. Cooper, C.~Murray, V.~Robert,
  G.~Tobie, T.~Guillot, S.~Mathis, F.~Remus, J.~Desmars, J.-E. Arlot, J.-P. {De
  Cuyper}, V.~Dehant, D.~Pascu, W.~Thuillot, C.~Le~Poncin-Lafitte, and J.-P.
  Zahn.
\newblock {New constraints on Saturn's interior from Cassini astrometric data}.
\newblock {\em Icarus}, 281:286, 2017.

\bibitem{Lopez13}
E.D. {Lopez} and J.J. {Fortney}.
\newblock {The Role of Core Mass in Controlling Evaporation: The Kepler Radius
  Distribution and the Kepler-36 Density Dichotomy}.
\newblock {\em ApJ}, 776:2, 2013.

\bibitem{Lopez18}
E.D. {Lopez} and K.~{Rice}.
\newblock {How formation time-scales affect the period dependence of the
  transition between rocky super-Earths and gaseous sub-Neptunes and
  implications for {\ensuremath{\eta}}$_{{\ensuremath{\oplus}}}$}.
\newblock {\em MNRAS}, 479:5303--5311, 2018.

\bibitem{Love11}
A.E.H. Love.
\newblock {\em {Some Problems of Geodynamics}}.
\newblock Cambridge University Press, 1911.
\newblock {Chap.~IV}.

\bibitem{Madhu20}
N.~{Madhusudhan}, M.C. {Nixon}, L.~{Welbanks}, A.A.A. {Piette}, and R.A.
  {Booth}.
\newblock {The Interior and Atmosphere of the Habitable-zone Exoplanet K2-18b}.
\newblock {\em ApJL}, 891:L7, 2020.

\bibitem{MayorQueloz95}
M.~{Mayor} and D.~{Queloz}.
\newblock {A Jupiter-mass companion to a solar-type star}.
\newblock {\em Nature}, 378:355--359, 1995.

\bibitem{Miller11}
N.~Miller and J.J. Fortney.
\newblock {The heavy-element masses of extrasolar giant planets, revealed}.
\newblock {\em ApJ}, 736:L29, 2011.

\bibitem{Morley17}
C.V. {Morley}, H.~{Knutson}, M.~{Line}, J.J. {Fortney}, D.~{Thorngren}, M.S.
  {Marley}, D.~{Teal}, and R.~{Lupu}.
\newblock {Forward and Inverse Modeling of the Emission and Transmission
  Spectrum of GJ 436b: Investigating Metal Enrichment, Tidal Heating, and
  Clouds}.
\newblock {\em AJ}, 153:86, 2017.

\bibitem{Mulders18}
G.D. {Mulders}.
\newblock {\em {Planet Populations as a Function of Stellar Properties. In:
  Handbook of Exoplanets.}}, page 153.
\newblock Eds.: {Deeg}, Hans J. and {Belmonte}, Juan Antonio, 2018.

\bibitem{Nettelmann19}
N.~Nettelmann.
\newblock {Tesseral harmonics of Jupiter from static tidal response}.
\newblock {\em ApJ}, 874:156, 2019.

\bibitem{Nettelmann11}
N.~Nettelmann, J.J. Fortney, U.~Kramm, and R.~Redmer.
\newblock {Thermal evolution and structure models of the transiting super-Earth
  GJ1214b}.
\newblock {\em ApJ}, 733:2, 2011.

\bibitem{Nettelmann10}
N.~Nettelmann, U.~Kramm, R.~Redmer, and R.~Neuh{\"a}user.
\newblock {Interior structure models of GJ436b}.
\newblock {\em A\&A}, 523:A26, 2010.

\bibitem{Neves12}
V.~{Neves}, X.~{Bonfils}, N.C. {Santos}, X.~{Delfosse}, T.~{Forveille},
  F.~{Allard}, C.~{Nat{\'a}rio}, C.~S. {Fernandes}, and S.~{Udry}.
\newblock {Metallicity of M dwarfs. II. A comparative study of photometric
  metallicity scales}.
\newblock {\em A{\&}A}, 538:A25, 2012.

\bibitem{OwenWu13}
J.E. {Owen} and Y.~{Wu}.
\newblock {Kepler Planets: A Tale of Evaporation}.
\newblock {\em ApJ}, 775:105, 2013.

\bibitem{Padovan18}
S.~{Padovan}, T.~{Spohn}, P.~{Baumeister}, N.~{Tosi}, D.~{Breuer}, Sz.
  {Csizmadia}, H.~{Hellard}, and F.~{Sohl}.
\newblock {Matrix-propagator approach to compute fluid Love numbers and
  applicability to extrasolar planets}.
\newblock {\em A{\&A}}, 620:A178, 2018.

\bibitem{Plotnykov20}
A.~{Plotnykov} and D.~Valencia.
\newblock {Chemical fingerprints of formation in rocky super-Earths' data}.
\newblock {\em MNRAS}, 499:932, 2020.

\bibitem{Poser19}
A.J. {Poser}, N.~{Nettelmann}, and R.~{Redmer}.
\newblock {The Effect of Clouds as an Additional Opacity Source on the Inferred
  Metallicity of Giant Exoplanets}.
\newblock {\em Atmosphere}, 10:664, 2019.

\bibitem{PuValencia17}
Bonan {Pu} and Diana {Valencia}.
\newblock {Ohmic Dissipation in Mini-Neptunes}.
\newblock {\em ApJ}, 846:47, 2017.

\bibitem{Scora20}
J.~{Scora}, D.~{Valencia}, A.~{Morbidelli}, and S.~{Jacobson}.
\newblock {Chemical diversity of super-Earths as a consequence of formation}.
\newblock {\em MNRAS}, 493:4910--4924, 2020.

\bibitem{Stevenson20}
David~J. {Stevenson}.
\newblock {Jupiter's Interior as Revealed by Juno}.
\newblock {\em Ann.~Rev.~EPS}, 48:465--489, 2020.

\bibitem{Madhu16}
S.W. {Thomas} and N.~{Madhusudhan}.
\newblock {In hot water: effects of temperature-dependent interiors on the
  radii of water-rich super-Earths}.
\newblock {\em MNRAS}, 458:1330--1344, 2016.

\bibitem{ThorngrenGao19}
D.~{Thorngren}, P.~{Gao}, and J.J. {Fortney}.
\newblock {The Intrinsic Temperature and Radiative-Convective Boundary Depth in
  the Atmospheres of Hot Jupiters}.
\newblock {\em ApJL}, 884:L6, 2019.

\bibitem{Thorngren18}
D.P. {Thorngren} and J.J. {Fortney}.
\newblock {Bayesian Analysis of Hot-Jupiter Radius Anomalies: Evidence for
  Ohmic Dissipation?}
\newblock {\em AJ}, 155:214, 2018.

\bibitem{Thorngren16}
D.P. Thorngren, J.J. Fortney, R.A. Murray-Clay, and E.D. Lopez.
\newblock {The Mass-Metallicity relation for giant planets}.
\newblock {\em ApJ}, 831:64, 2016.

\bibitem{Turbet19}
M.~{Turbet}, D.~{Ehrenreich}, C.~{Lovis}, E.~{Bolmont}, and T.~{Fauchez}.
\newblock {The runaway greenhouse radius inflation effect. An observational
  diagnostic to probe water on Earth-sized planets and test the habitable zone
  concept}.
\newblock {\em A\&A}, 628:A12, 2019.

\bibitem{Valencia13}
D.~Valencia, T.~Guillot, V.~Parmentier, and R.S. Freedman.
\newblock {Bulk composition of GJ1214b and other sub-Neptune exoplanets}.
\newblock {\em ApJ}, 775:10, 2013.

\bibitem{Vazan18}
A.~{Vazan}, R.~{Helled}, and T.~{Guillot}.
\newblock {Jupiter's evolution with primordial composition gradients}.
\newblock {\em A{\&}}, 610:L14, 2018.

\bibitem{Venturini16}
J.~Venturini, Y.~Alibert, and W.~Benz.
\newblock {Planet formation with envelope enrichment: new insights on planetary
  diversity}.
\newblock {\em A{\&}A}, 596:90, 2016.

\bibitem{Wahl20}
S.M. {Wahl}, M.~{Parisi}, W.M. {Folkner}, W.B. {Hubbard}, and B.~{Militzer}.
\newblock {Equilibrium Tidal Response of Jupiter: Detectability by the Juno
  Spacecraft}.
\newblock {\em ApJ}, 891:42, 2020.

\bibitem{Wahl13}
S.M. {Wahl}, H.F. {Wilson}, and B.~{Militzer}.
\newblock {Solubility of Iron in Metallic Hydrogen and Stability of Dense Cores
  in Giant Planets}.
\newblock {\em ApJ}, 773:95, 2013.

\bibitem{Wakeford18}
H.R. {Wakeford}, D.K. {Sing}, D.~{Deming}, N.K. {Lewis}, J.~{Goyal}, T.J.
  {Wilson}, J.~{Barstow}, T.~{Kataria}, B.~{Drummond}, T.M. {Evans}, A.L.
  {Carter}, N.~{Nikolov}, H.A. {Knutson}, G.E. {Ballester}, and A.~M.
  {Mandell}.
\newblock {The Complete Transmission Spectrum of WASP-39b with a Precise Water
  Constraint}.
\newblock {\em AJ}, 155:29, 2018.

\bibitem{Zeng19}
L.~{Zeng}, S.B. {Jacobsen}, D.D. {Sasselov}, M.I. {Petaev}, A.~{Vanderburg},
  M.~{Lopez-Morales}, J.~{Perez-Mercader}, T.R. {Mattsson}, G.~{Li}, M.Z.
  {Heising}, A.S. {Bonomo}, M.~{Damasso}, T.A. {Berger}, H.~{Cao}, A.~{Levi},
  and R.D. {Wordsworth}.
\newblock {Growth model interpretation of planet size distribution}.
\newblock {\em PNAS}, 116:9723--9728, 2019.

\bibitem{Zhang20}
X.~Zhang.
\newblock {Atmospheric regimes and trends on exoplanets and brown dwarfs}.
\newblock {\em Res.~Astron.~Astroph.}, 20:99, 2020.

\end{thebibliography}


\end{document}